# High Performance Three-Terminal Thyristor RAM with a P$^+$/P/N/P/N/N$^+$ Doping Profile on a Silicon-Photonic CMOS Platform

Changseob Lee[1], Ikhyeon Kwon, [1], Anirban Samanta[1], Siwei Li[1], and S. J. Ben Yoo[1]
[1]University of California, Davis, CA, USA, email: kcslee@ucdavis.edu

*Abstract*—This work demonstrates the experimental implementation of Three Terminal Thyristor RAM (3T TRAM) using a newly proposed doping structure (P$^+$PNPNN$^+$), on a silicon photonic platform. By using additional implant layers, this device provides excellent memory performance compared to the conventional structure (P$^+$NPN$^+$) fabricated on the same platform. Specifically, the memory window increases by 1V, the ON/OFF ratio improves by ×10$^2$, and the holding current is reduced by ×10$^{-5}$. A physics-based model is required to predict the advanced design of this device. Therefore, we detail how the TCAD modeling process is used to reflect the physical behavior, and the high-speed memory operations are described through a hardware-reflected model. We investigate the memory operation using the modeled structure, showing that program voltage of 0.8 V and holding voltage of 0.5 V, which can potentially be integrated with <1V CMOS technology [1].

## I. Introduction

Emerging applications such as training large machine learning models, performing graph analytics, and utilizing large language models (LLMs) require memory capacities in the tens of terabytes, with computational demands growing exponentially. However, traditional memory technologies such as static random-access memory (SRAM), dynamic random-access memory (DRAM), and flash memory are unable to fulfill these requirements [2, 3]. DRAMs face challenges in monolithic integration with CMOS-based processors (CPUs/GPUs) due to their deep-trench capacitor design, resulting in larger cell sizes and the need for frequent refresh operations, making them incompatible with the newly emerging 2-3 nm technology nodes. There is no suitable existing memory technology that supports < 10 F$^2$ cell area, < 2 ns read/write latency available today, especially for below 10 nm technology node, except for one technology: Thyristor RAM (TRAM). In recent times, TRAMs have quickly gained attention as a potential future memory technology compatible with upcoming 3 nm to 2 nm Gate-All-Around FET (GAA-FET) and Complementary FET (CFET) technologies [4], [5]. Today's 2.5D and 3D electronic interconnects and packaging are rapidly advancing to mitigate some of these challenges; however, they still suffer from fundamental limitations of electrical wire impedances, electromagnetic-interferences, and pin densities.

To overcome these, recent advancements in silicon photonics, quantum-impedance conversion through close integration of photonics and electronics [6], and 3D electronic-photonic integrated circuits (3D EPICs) present promising avenues for future computing [7], [8]. These innovations offer potential improvements in scalability, performance, and energy efficiency when new materials and devices are co-designed and co-integrated with new computing system architectures [9], driven by modern application requirements.

With this objective, we fabricated TRAM using a conventional structure on a silicon photonic platform to facilitate electronic-photonic co-design. In this work, we enhanced memory performance by incorporating additional doping layers. Furthermore, we model the memory operations using a physics-based TCAD model to enable predictive design, facilitating integration with matured CMOS compact models.

## II. Experimental and Modeling

### A. Model Hardware Correlation (MHC)

The correlation between the hardware and the physics-based model is crucial for validating the reliability of emerging devices and enables us to design optimized structures. Fig. 1 shows the modeled $I_a$ (anode current)-$V_a$ (anode voltage) characteristics of newly fabricated 3T TRAM with P$^+$/P/N/P/N/N$^+$ structure compared to previously studied P$^+$/N/P/N$^+$ structure [10]. We experimentally demonstrate a significant improvement in memory performance with the additional doping structure. Specifically, the memory window increased by 1V. The sensing margin (ON/OFF ratio) improved by ×10$^2$. Also, a holding current ($I_{hold}$) is reduced by ×10$^{-5}$, reducing the power consumption.

### B. Device Structure and Measurement

Fig. 2 illustrates the detailed implant layers and the fabricated device structure. The 3T TRAM is implemented on a monolithic silicon photonic platform using GlobalFoundries' 45nm CMOS technology. The P$^+$/P/N/P/N/N$^+$ layers are designed with a length of 200/400/200/400/400/200 nm, respectively, and a width of 150 nm. The P$^+$/N/P/N$^+$ layers are designed with a length of 300/200/400/300 nm, respectively, and a width of 200 nm. The thickness of the device and SiO$_2$ layers are manufactured to be 200 nm and 5 nm, respectively. Geometries of devices are designed following the Design Rule Check (DRC) guidelines provided with the Process Design Kit (PDK), which include specifications such as minimum spacing, minimum area of layers, and other critical dimensions. The measurement setup, test chips, and PAD images are illustrated in Fig. 3. The designed devices are fabricated alongside other devices intended for different applications in a multi-project wafer (MPW). The PADs are designed to be compatible with a probe station through the metals.

### C. Physical Principle and Modeling Methodology

Physical principles such as the band theory and Poisson's equations are considered to describe the behavior of the device and to design its structure. Fig. 4(a) illustrates the energy band transitions during the rising of anode voltage ($V_a$). From an equilibrium state (0%) of the rising, the energy band on the *p*$^+$ side is lowered by the injected hole carriers from the anode, and the device rapidly (80%-90%) turns on once all the barrier heights become flattened. In the falling process (Fig. 4(b)), the energy band returns to the equilibrium state. Unlike the rising process, all the bands throughout the device move up evenly due to the uniformly distributed voltage drop caused by the low impedance junctions of the already turned-on device. The device turns off gradually (60%-100%), providing candidates for a holding voltage ($V_{hold}$), which can return to the turn-on state by a read voltage. Carriers are stored inside during the holding state and placed inside a moderately bent energy band. The device distinguishes between 'Read1' and 'Read0' based on whether the carriers flow out during the read operation.

In order to optimize this process and design the structure, the electric field ($E_{field}$) is investigated in Fig. 5, highlighting the contribution of the junctions. It shows the $E_{field}$ during the rising process: before Junction 1 ($J_1$) turns on (0-40%), after it turns on

(50%-80%), and at full turn-on. The $J_1$ contributes to the transition of the $E_{field}$ until it turns on, and a portion of this contribution can be increased by the lightly doped p-type slab layer. At the initial stage, $V_a$ is distributed not only to $J_1$ but also to $J_2$, which has high impedance due to reverse bias. This delays the turn-on point and allows for the adjustment of the memory window. The increased memory window enhances memory reliability without requiring a high program voltage ($V_{pg}$), as $V_{pg}$ can be effectively reduced by the gate voltage ($V_g$). Fig. 6 shows that this doping profile also helps reduce the $V_{hold}$ during the falling process. The $E_{field}$ transition moves evenly during the low impedance stage, whereas once the turn-off process starts, the transition concentrates on the junctions. By placing the n-type slab layer in contact with the $n^{++}$ photonic doping, the transition in $J_3$ can be balanced with those in $J_1$ and $J_2$. This prevents an abrupt turn-off by lowering the subthreshold swing (SS) and extends the high impedance region, enabling the selection of a stable $V_{hold}$.

Taking advantage of the combination of doping profiles, the side of the structure should be highly doped to operate as a memory device, allowing for deep energy band bending where carriers can be charged. Fig. 7 shows the dependency of on-current ($I_{on}$) on the variation in doping concentration of the implant layers. Once the device turns on, all the junctions created by the contact of PN layers switch to forward bias, resulting in an almost flat energy band. $I_{on}$ is negligibly affected by the variation in doping concentration of the inside layers, as it is primarily influenced by the highly doped layer on the side. By maintaining high doping concentrations on both sides and lowering the Schottky barrier with the electrodes, the ON/OFF ratio can be preserved. Fig. 8 shows that, in summary, the memory performance is enhanced through an additional doping profile while maintaining the intrinsic properties of the memory device. The doping concentration of the p-type body under the gate is provided by GlobalFoundries and is specified as $10^{16}$ cm$^{-3}$. To create a balanced junction, the MZM p/n-type slab layers are added, and the existing memory performance is maintained with the relatively high MZM n-type rib layer doping and heavily doped photonic P and $N^{++}$ layers on both sides.

### III. Memory Operation

The gate terminal of the 3T TRAM adjusts the energy band, allowing for fine-tuning of the memory characteristics [11]. Here, we use a modeled structure in Sentaurus TCAD to describe the $V_{pg}$, and $V_{hold}$, and high-speed memory characteristics with respect to $V_g$. In Fig. 9(a), during the program process, a positive $V_g$ is applied to lower the energy band of the p-body, allowing the device to temporarily experience $V_{pg}$. During the holding process, $V_a$ is maintained at $V_{hold}$, and $V_g$ is returned to 0, increasing the memory window and storing carriers in the device. During the read operation, $V_a$ is set to $V_{read}$ to sense the presence of carriers in the device and determine a 0 or 1. Subsequently, the device can be erased by reducing $V_a$. Fig. 10 shows the $I_a$-$V_g$ plot corresponding to different $V_a$ values. For example, in the curve where $V_a=1$, $V_g$ increases, which decreases $V_{pg}$, enabling programming. Unlike when sweeping $V_a$, even if $V_g$ returns to 0, $I_a$ is maintained within a wide memory window as long as $V_{hold}$ is sustained. This allows determining $V_{hold}$ to minimize $I_a$.

Fig. 11 compares the scenarios with and without $V_g$, showing how much $V_{pg}$ can be reduced. In the initial scenario without $V_g$ (Fig. 11(a)), only when $V_a$ exceeds $V_{pg}$ is sensed during the read operation (Fig. 11(b)). However, with $V_g$ applied at only 0.5V (Fig. 11(c)), $V_a$ can be programmed at 0.8V, and even if $V_{hold}$ is maintained at 0.5 V, the current is sensed (Fig. 11(d)). However, as shown in the figure, if $V_{hold}$ is not enough to maintain the charge, a collapsed $I_{sense}$ pulse can be observed, as described in Fig. 12. Raising $V_{hold}$ to 0.6 V can achieve a stable $I_{sense}$ without collapse, however $I_{hold}$ increases to the 0.3 nA level. Conversely, lowering $V_{hold}$ causes $I_{sense}$ to collapse. However, it allows $I_{hold}$ to be reduced to the 0.1 pA level at $V_{hold}$ of 0.4 V. This trade-off can be utilized to meet the design requirements for speed and power consumption, necessitating an optimal $V_{hold}$ setting.

Therefore, at $V_{hold}$=0.6 V, which ensures read reliability, the speed limitation with and without $V_g$ is investigated in Fig. 13. When $V_g$ is not applied, $V_{pulse}$ is not sensed starting from 200 ns (Fig. 13(a)). On the other hand, when $V_g$ is programmed together at 0.5V, programming is enabled up to 300 ps, and speed limitations occur at 200 ps (Fig. 13(b)). However, even if programming occurs with a 400 ps $T_{pulse}$, factors that deteriorate reliability emerge at the high-speed operation. Fig. 14 shows that in the timing diagram, high-speed falling is not used to set timing limitations while inducing reverse bias. This reverse current peaks at the mA level and can readily degrade the device. This effect is observed when a pulse, which is too fast to be programmed during the $T_{rise}$ speed but slow enough to be charged within $T_{hold}$, falls, as shown in Fig. 14(b). Fig. 15 shows the distribution of the $E_{field}$ when $V_a$ falls at high speed and low speed. During the initial falling period, the device is turned on with low impedance, causing the $E_{field}$ transition to correspond to the high speed $V_a$, similar to the low speed observed earlier. However, once the turn-off begins, the devices cannot instantaneously respond to changes in $V_a$ beyond the depletion capacitance. For high-speed design, these factors must be considered along with the $T_{rise}$ portion, which plays a dominant role in the program time [12].

### IV. Conclusion

The 3T TRAM with a $P^+/P/N/P/N/N^+$ structure is successfully developed, demonstrating an increased memory window, improved sensing margin, and reduced holding voltage on a silicon-photonic platform. The doping profile was designed using a combination of implant layers currently used in silicon-photonic design without additional fabrication steps. Furthermore, the structure is modeled using the TCAD simulation through the physical analysis, and the dynamic characteristics are thoroughly analyzed using modeled simulations. The implementation of high-performance memory devices using photonic layers provided by the PDK offers significant potential for advancement through co-design with emerging photonic design methodologies [13]. This approach enables co-design in 3D EPIC designs and paves the way for future in-memory computing technologies.

### Acknowledgment

This work was supported in part by ARO award W911NF1910470.

### References

[1] T.-H. Hsu et al., IEEE J. Solid-State Circuits PP, 1 (2020).
[2] R. Roy et al., ISSCC (2006), 2612.
[3] M. Bohr, IEEE Trans. Nanotechnol. 1, 56 (2002).
[4] K. Lee et al., IEDM (2023), 1.
[5] W.-C. Chen et al., IEDM (2023), 1.
[6] D. A. Miller, Opt. Lett. 14, 146 (1989).
[7] S. Werner et al., Int. Symp. Memory Syst. (2019), 206.
[8] S. B. Yoo, SPIE 10926, 109261T (2019).
[9] P. Fotouhi et al., Int. Conf. High Perform. Comput. (2021), 176.
[10] I. Kwon et al., CLEO (2024), Optica Publishing Group.
[11] D. Lim et al., Adv. Mater. Technol. 5, 2000915 (2020).
[12] B. -S. Lee et al., Nanotechnology 32, 14LT01 (2021).
[13] R. A. de Paula Jr. et al., Nature, 13, 14662 (2023)

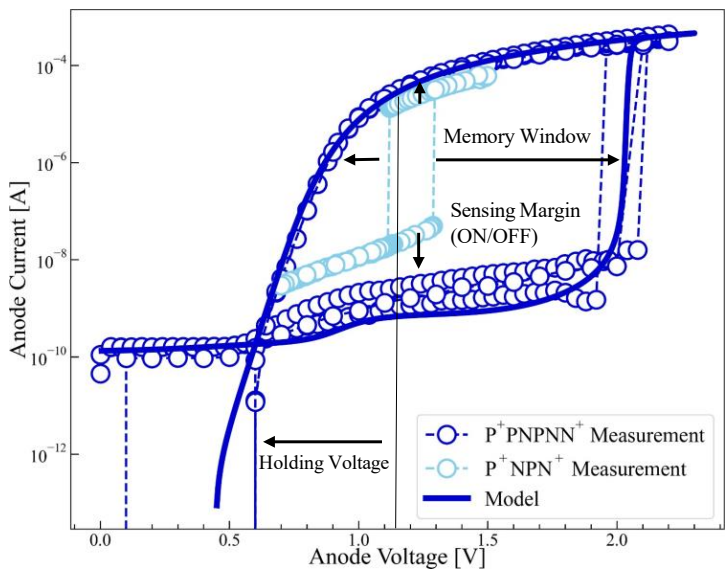

Fig. 1. Modeled $I_a$-$V_a$ characteristics of 3T TRAM with fabricated measurement of $P^+/P/N/P/N/N^+$ structure. Comparison to conventional $P^+/N/P/N^+$ structure.

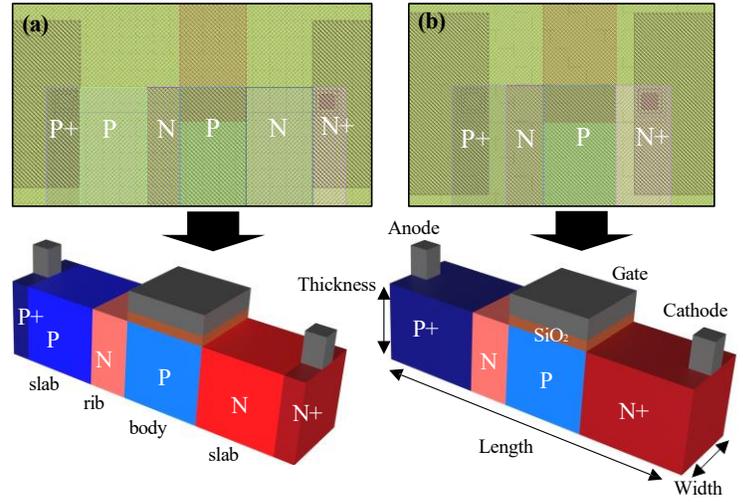

Fig. 2. (a) Implant layers of photonic $P^{++}$, MZM p-type slab, MZM n-type rib, MZM p-type body, MZM n-type slab, and photonic $n^{++}$ are used in $P^+/P/N/P/N/N^+$ structure. (b) Photonic $P^{++}$, MZM n-type rib, MZM p-type body, photonic $n^{++}$ in $P^+/N/P/N^+$ structure.

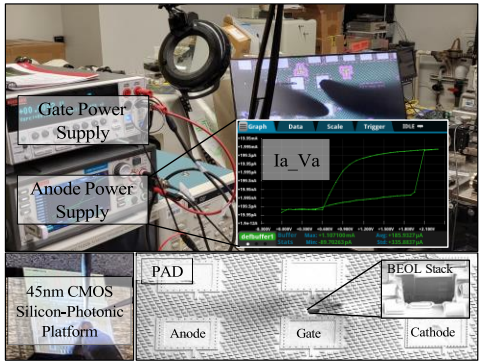

Fig. 3. Keithley 2450 and Keithley 2401 are used to apply anode and gate bias, respectively. The PAD image includes SEM (outer) and TEM (inner) views.

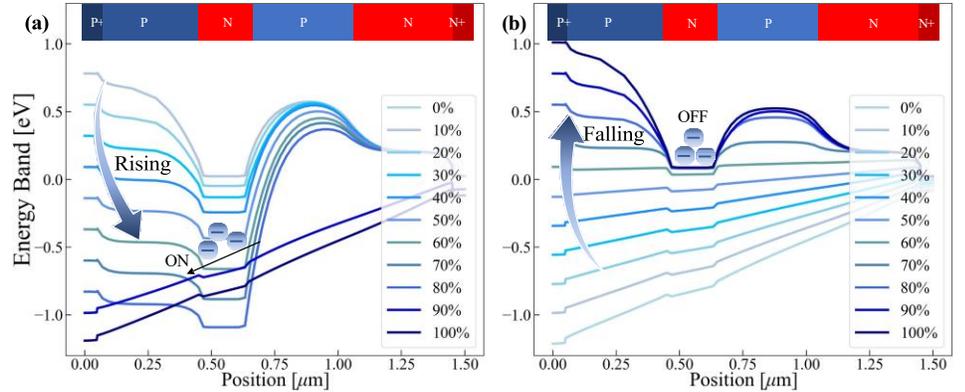

Fig. 4. Energy transition during the rising and falling of the anode voltage. The plots are extracted by cutting the modeled TCAD from the $P^{++}$ photonic part (left) to the $N^{++}$ photonic part (part).

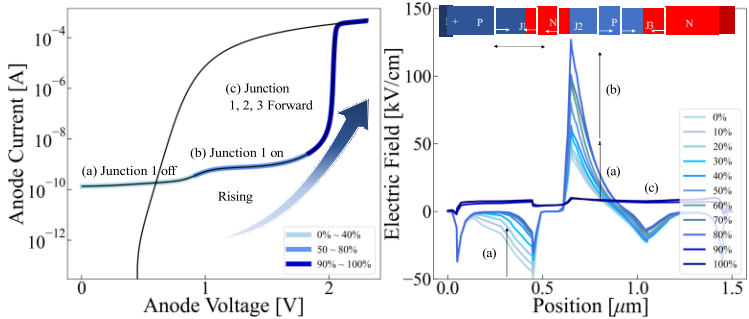

Fig. 5. Electric field during the rising process before (0-40%) and after (50-80%) the junction 1 turns on, and when all junctions are turned on (80-100%).

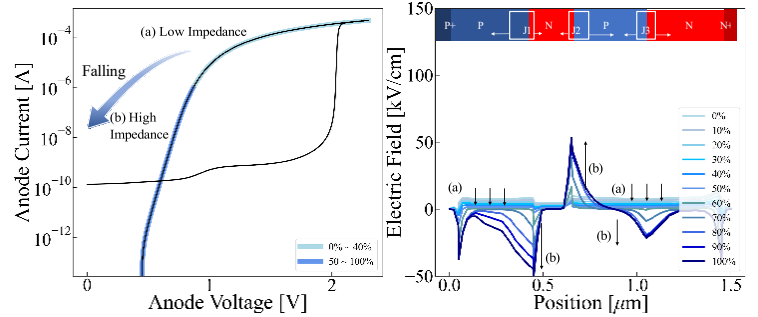

Fig. 6. Electric field during the falling process before (0-40%) and after (50-100%) the device starts to turn off.

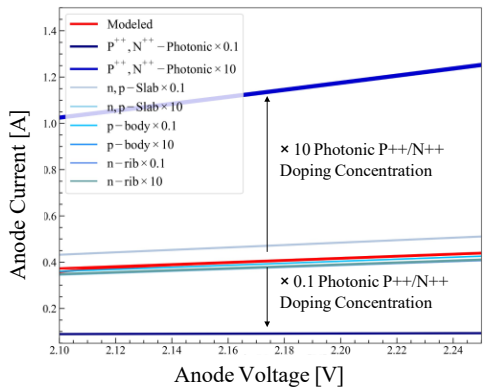

Fig. 7. Dependency of the $I_{on}$ on the variation in doping concentration of the doping layers.

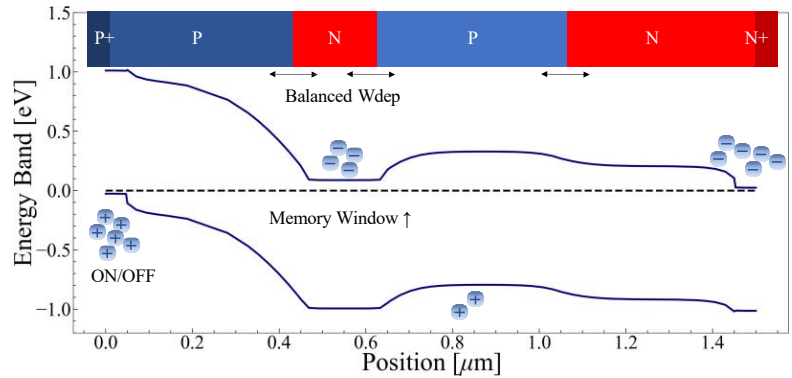

Fig. 8. Energy band diagram illustrating enhanced memory performance with additional doping profile and balanced junctions

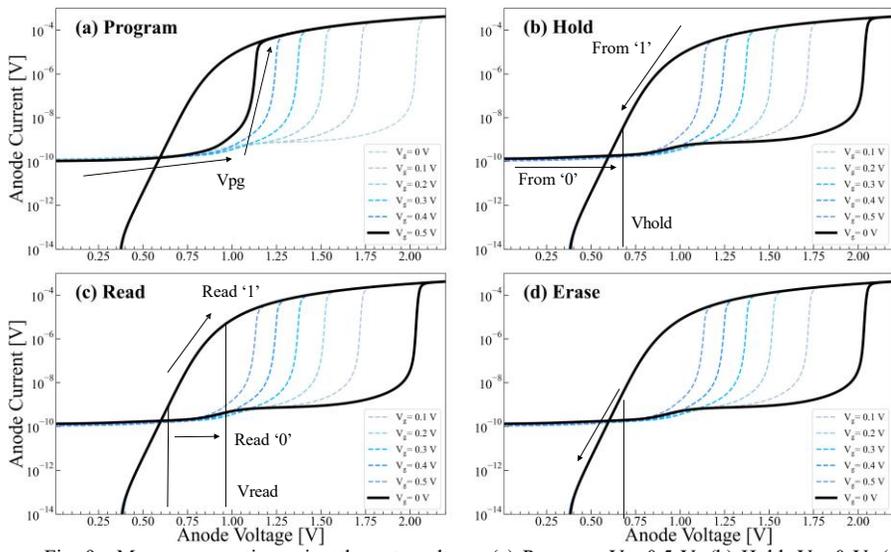
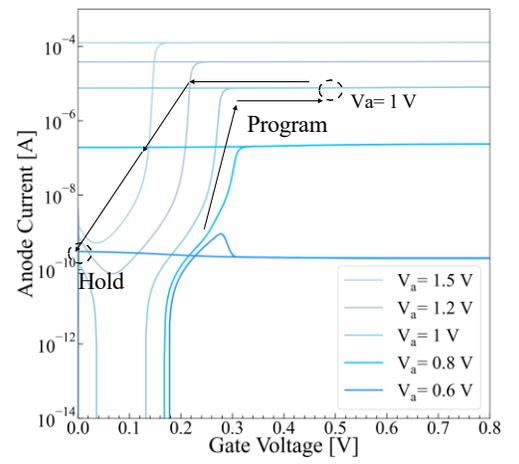

Fig. 9. Memory operation using the gate voltage. (a) Program: $V_g$=0.5 V, (b) Hold: $V_g$=0 V, (c) Read: $V_g$=0 V, (d) Erase: $V_g$=0 V.

Fig. 10. $I_a$-$V_g$ characteristics of 3T-TRAM with different anode voltages.

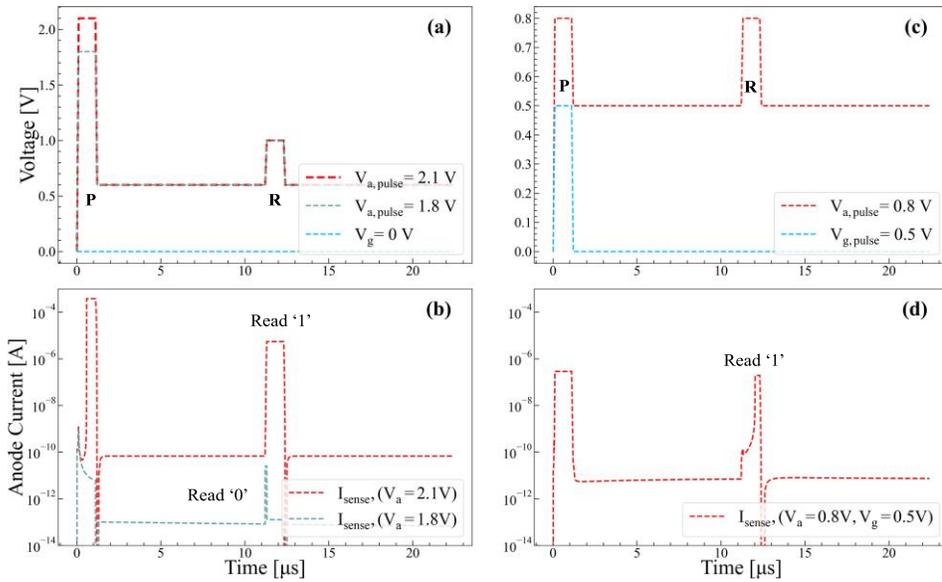
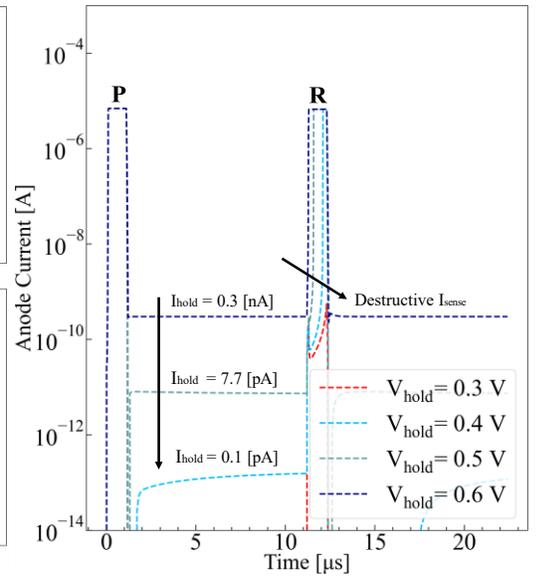

Fig. 11. Program and read operations without $V_g$: (a) $V_{pulse}$, (b) $I_{sense}$; and with $V_g$: (c) $V_{pulse}$, (d) $I_{sense}$ ($T_{pulse}$=1 us, $T_{rise,fall}$=0.1 us, $T_{hold}$=10 us).

Fig. 12. Program and read operations as a function of the $V_{hold}$ ($T_{pulse}$=1 us, $T_{rise,fall}$=0.1 us, $T_{hold}$=10 us).

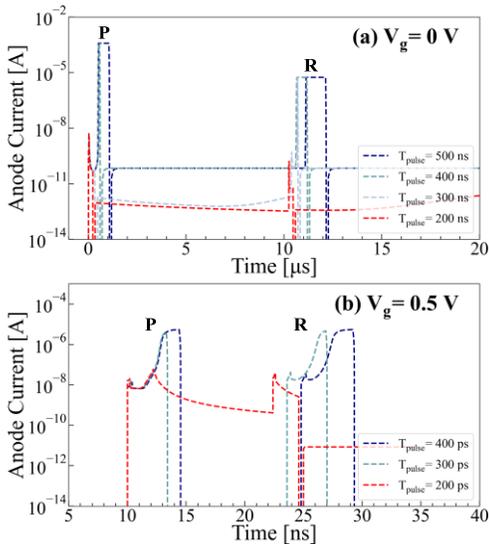
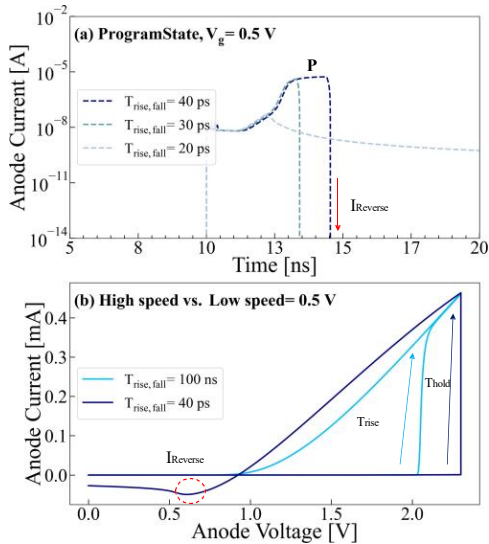
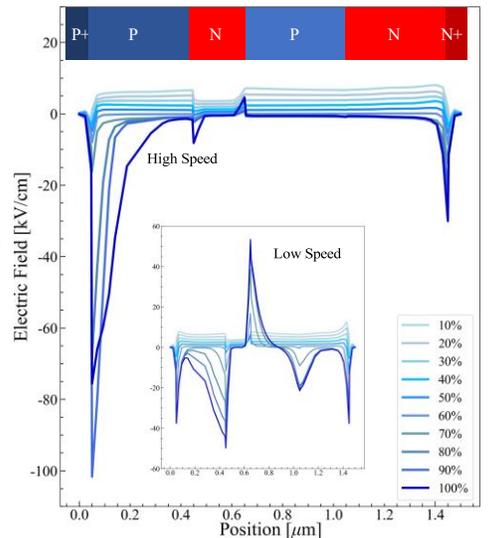

Fig. 13. Speed limitation: (a) without $V_g$: $V_{pg}$=2.1 V, $V_{hold}$=0.6 V, $V_{read}$=1 V; (b) with $V_g$: $V_{pg}$=1 V, $V_{hold}$=0.6 V, $V_{read}$=1 V ($T_{rise,fall}$=$T_{pulse}$×0.1, $T_{hold}$=$T_{pulse}$ × 10).

Fig. 14. High-speed program process $I_a$-time and $I_a$-$V_a$ characteristics.

Fig. 15. $E_{field}$ distribution at a high-speed and low-speed falling.